\documentclass[11pt]{article}
\usepackage{aaspp4}
\usepackage{graphics}

\makeatletter

\renewcommand{\maketitle}{}
\newcommand{\bmath}[1]{\mbox{\boldmath\(#1\)}}

\makeatother

\begin{document}

\slugcomment{Submitted to ApJ}
\newcommand{\dW}{\Delta \widetilde{\Omega }}

\newcommand{\dWm}{\delta \widetilde{\Omega }}

\newcommand{\dL}{\Delta \widetilde{L}}

\newcommand{\dE}{\Delta \widetilde{E}}

\newcommand{\rp}{\widetilde{r}_{p}}

\newcommand{\Wp}{\widetilde{\Omega }_{p}}

\newcommand{\mi}{\widetilde{m}}

\newcommand{\si}{\widetilde{\sigma }}

\newcommand{\MOI}{\widetilde{I}}

\newcommand{\Eo}{\widetilde{E}_{o}}

\newcommand{\Eb}{\widetilde{E}_{b}}

\newcommand{\mr}{\widetilde{\mu }}
 
\newcommand{\Tt}{\widetilde{T}}

\newcommand{\vi}{\widetilde{v}_{\infty }}

\newcommand{\vp}{\widetilde{v}_{p}}

\newcommand{\ns}{\widetilde{n}}

\newcommand{\rt}{\widetilde{r}_{t}}

\newcommand{\rc}{\widetilde{r}_{c}}

\newcommand{\qc}{\widetilde{q}}

\title{Tidal spin-up of stars in dense stellar cusps around massive black holes }

\author{Tal Alexander \& Pawan Kumar}

\author{\affil{Institute for Advanced Study, Olden Lane, Princeton, NJ 08540} }

\maketitle
\begin{abstract}
We show that main-sequence stars in dense stellar cusps around massive black
holes are likely to rotate at a significant fraction of the centrifugal breakup
velocity due to spin-up by hyperbolic tidal encounters. We use realistic stellar
structure models to calculate analytically the tidal spin-up in soft encounters,
and extrapolate these results to close and penetrating collisions using smoothed
particle hydrodynamics simulations. We find that the spin-up falls off only
slowly with distance from the black hole because the increased tidal coupling
in slower collisions at larger distances compensates for the decrease in the
stellar density. We apply our results to the stars near the massive black hole
in the Galactic Center. Over their lifetime, \( \sim \! 1\, M_{\odot } \) main
sequence stars in the inner 0.3 pc of the Galactic Center are spun-up on average
to \( \sim \! 10\% \)--\( 30\% \) of the centrifugal breakup limit. Such rotation
is \( \sim \! 20 \)--\( 60 \) times higher than is usual for such stars and
may affect their subsequent evolution and their observed properties. 
\end{abstract}

\section{Introduction}

\label{sec: intro}

It is now widely accepted that super-massive black holes (BHs) exist in many,
if not all galactic centers (Magorrian et al. \cite{Mag98})\@. Dynamical models
of the evolution of such systems generically predict the formation of a dense
stellar cusp near the BH (e.g. Bahcall \& Wolf \cite{Bah77}; Young \cite{You80}).
Stars moving rapidly in the dense stellar cusp near the BH will suffer numerous
high-velocity close tidal encounters over their lifetimes. Although such encounters
transfer some energy and angular momentum from the hyperbolic orbit to the colliding
stars, they rarely remove enough energy for tidal capture. This is in marked
contrast to the situation in the high density cores of globular clusters, where
the colliding stars are on nearly zero-energy orbits and close collisions lead
to the formation of tight binaries. The effects of hyperbolic encounters on
the stars are mostly transient. The dynamical and thermal relaxation timescales
are very short compared to the stellar lifetime, and thus apart from some mass-loss
in very close collisions, the star is largely unaffected. It is however more
difficult for the star to shed the excess angular momentum since magnetic breaking
operates on timescales of the order of the stellar lifetime (Gray \cite{Gra92}).
High rotation is therefore the longest lasting dynamical after-effect of a close
encounter. 

The possibility that stars in BH cusps are rapid rotators may have interesting
implications for their evolution and the interpretation of their observed properties.
The effects of rotation and rotationally induced mixing in a main-sequence (MS)
star on its subsequent evolution has been studied by Sweigart (\cite{Swe97})
and VandenBerg, Larson \& De Propris (\cite{Van98}) in the context of globular
clusters, without specifying the origin of the rotation, and by Sills, Pinsonneault
\& Terndrup (\cite{Sil00}) in the context of young open clusters. Rotationally
induced mixing may reveal itself in the spectral line ratios, and rotation may
be directly observed in the spectral line profiles. Detection of such signatures
in the spectra of the observed giants can provide additional evidence for the
existence of an underlying cusp of MS stars, which at present cannot be directly
observed.

The goal of this study is to estimate the magnitude of the tidal spin-up, with
particular emphasis on the Galactic Center (GC). Present-day observations can
already resolve individual giant stars very close to the BH in the GC (Genzel
et al. \cite{Gen97}; Ghez \cite{Ghe98}) and high resolution infrared spectroscopy
is possible for the brighter giants (Carr, Sellgren \& Balachandran \cite{Car99};
Ramirez et al. \cite{Ram00}). The depletion of luminous giants in the inner
\( 2'' \) (\( \sim \! 0.1 \) pc) around the supermassive BH in the Galactic
Center (GC) was interpreted by Alexander (\cite{Ale99a}) as evidence for collisional
destruction in an extreme density of a sharp stellar cusp. It is inevitable
that where the stellar density is high enough to destroy giants, smaller stars
that escape destruction will suffer very close collisions. The inner GC is therefore
a promising environment for studying the spin-up effect.

The paper is organized as follows. In Sec.~\ref{sec: soft} we present the formalism
for spin-up in the linear regime of soft hyperbolic encounters. In Sec.~\ref{sec: strong}
we discuss results from SPH simulations of non-linear close encounters and incorporate
them in our spin-up calculations. In Sec.~\ref{sec: results} we calculate the
spin-up of stars in the inner parsec of the GC. We discuss and summarize our
results in Sec.~\ref{sec: discuss}. An appendix describes the analytic calculation
of the tidal coupling constants for hyperbolic encounters using realistic stellar
structure models of a MS dwarf and of a giant.

\section{Spin-up by soft hyperbolic tidal encounters}

\label{sec: soft}

\subsection{Stochastic spin-up in the linear regime}

We begin by considering soft (distant) encounters where the tidal deformations
are small enough to be treated as linear perturbations. We consider the effect
of the tides raised by an impactor star of mass \( m \) on a target star of
mass \( M \) and radius \( R \) as the impactor follows an unbound orbit with
a peri-separation \( r_{p} \) from the target star. We will use the tilde symbol
to denote mass in terms of \( M \), distances in terms of \( R \), time in
terms of \( \sqrt{R^{3}/GM} \), velocity in terms of the Keplerian velocity
\( \sqrt{GM/R} \), energy in terms of \( GM^{2}/R \), angular momentum in
terms of \( \sqrt{GM^{3}R} \) and moment of inertia in terms of \( MR^{2} \).
In these units \( \widetilde{\Omega }=1 \) is the centrifugal break-up angular
frequency. A star with \( \widetilde{\Omega }>1 \) will shed mass from its
equator due to the centrifugal force.

We describe the results in the reduced mass system where the target star is
at the origin. The angular momentum \( \dL  \) that is transferred from the
orbit to the target star is related to the deposited tidal energy \( \dE  \)
by (Goldreich \& Nicholson \cite{Gol89}; Kumar \& Quataert \cite{Kum98})
\begin{equation}
\dE =\dL \Wp \, ,
\end{equation}
where \( \widetilde{\Omega }_{p} \) is the angular velocity at periastron.
The energy invested in raising the tides is given in the linear regime by (Press
\& Teukolsky \cite{Pre77})
\begin{equation}
\dE =\frac{\widetilde{m}^{2}}{\rp ^{2}}\sum ^{\infty }_{l=2}\frac{T_{l}(\eta )}{\rp ^{2l}}\, ,
\end{equation}
where the tidal coupling coefficients \( T_{l} \) depend on the structure of
the star and the eccentricity of the orbit and are functions of the parameter
\( \eta  \),
\begin{equation}
\eta =\sqrt{\frac{\rp ^{3}}{1+\mi }}\, .
\end{equation}
 For rigid body rotation\footnote{%
The timescale for angular momentum re-distribution due to convective transport
in a late type giant is \( \sim \! 1 \) yr (Zahn \cite{Zah89}) and so rigid
rotation is achieved on a timescale similar to that of the collision itself.
The timescale for angular momentum re-distribution in radiative MS stars is
not well known, although it is likely to be shorter than the stellar lifetime. 
}, the change in the star's angular momentum is related to the change in the
angular velocity \( \dW  \) by the star's moment of inertia \( \MOI  \),
\begin{equation}
\label{eq:rigidL}
\dL =\MOI \dW \, .
\end{equation}
In the case of differential rotation, \( \dW  \) is defined by Eq.~\ref{eq:rigidL}
and is the effective angular velocity. The spin-up of the target star in a single
tidal encounter can therefore be expressed as
\begin{equation}
\label{eq:dwt}
\dW =\frac{\mi ^{2}}{\MOI \Wp \rp ^{2}}\sum ^{\infty }_{l=2}\frac{T_{l}(\eta )}{\rp ^{2l}}\, ,
\end{equation}
where it is assumed that the star maintains its original moment of inertia.
The periastron angular velocity is related to the relative velocity at infinity,
\( \widetilde{v}_{\infty } \) by 
\begin{equation}
\label{eq:wp}
\Wp ^{2}=\frac{\vi ^{2}}{\rp ^{2}}+\frac{2(1+\mi )}{\rp ^{3}}\, ,
\end{equation}
where the second term expresses the enhancement due to gravitational focusing. 

Many encounters at random orientations lead to a ``random walk'' buildup of
the stellar spin. Over the stellar lifetime \( \widetilde{T} \), the rms change
in the stellar angular velocity is given by adding in quadrature the contributions
from collisions with different values of the peri-separation \( \rp  \) and
of the orbital energy \( \widetilde{E}_{o} \), weighted by the differential
collision rate \( d^{2}\qc /d\rp d\Eo  \),

\begin{equation}
\label{eq:edw}
\delta \widetilde{\Omega }\equiv \left\langle \dW ^{2}\right\rangle ^{1/2}=\left( \Tt \int d\rp \int d\Eo \dW ^{2}\frac{d^{2}\qc }{d\rp d\Eo }\right) ^{1/2}\, .
\end{equation}
The differential rate is calculated with the approximation that the relative
velocity can be described by the Maxwellian distribution function (DF) with
a mass-independent 1D velocity dispersion \( \widetilde{\sigma }_{2}^{2}=2\widetilde{\sigma }^{2} \)
(Sec. \ref{sec:vDF}) 
\begin{equation}
\label{eq:dq}
\frac{d^{2}\qc }{d\rp d\Eo }=\frac{\sqrt{8\pi }\ns }{\mr ^{2}\widetilde{\sigma }^{3}_{2}}\exp \left( -\frac{\Eo }{\mr \widetilde{\sigma }_{2}^{2}}\right) \left( 2\rp \Eo +\mi \right) \, ,
\end{equation}
where \( \ns  \) is the space density of the impactors. This estimate involves
the approximations that the angular momentum transfer does not depend on whether
the star is rotating in a prograde or retrograde sense with respect to the orbit,
and it assumes that the mass-loss and the change in the star's structure and
moment of inertia can be neglected. Because \( \dW ^{2}\sim \rp ^{-9} \) whereas
\( d^{2}\qc /d\rp d\Eo \sim \rp  \), the rms \( \delta \widetilde{\Omega } \)
is dominated by the collisions with the smallest \( \rp  \). When the stellar
population includes a spectrum of masses, \( \int f_{\mi }d\widetilde{m}=1 \),
the average spin-up is obtained by adding the weighted contributions in quadrature
\begin{equation}
\label{eq:edwm}
\dWm \equiv \left\langle \int d\mi f_{\mi }\dW ^{2}\right\rangle ^{1/2}\, .
\end{equation}

\subsection{Linear tidal coupling coefficients }

\label{sec:lin}

The orbital energy \( \Eo  \) is related to \( \vi  \) and the eccentricity
of the orbit \( e \) by 
\begin{equation}
\Eo =\frac{1}{2}\mr \vi ^{2}=\frac{1}{2}\mr \frac{e-1}{\rp }\, ,
\end{equation}
 where \( \mr =\mi /(1+\mi ) \) is the reduced mass. The eccentricity that
corresponds to the mean orbital energy (Eq.~\ref{eq:avEo} below) is
\begin{equation}
\left\langle e\right\rangle =6\rp \si ^{2}+1\, .
\end{equation}
Since \( e \) can reach very high values when \( \si \gg 1 \), as is the case
for giants (low Keplerian velocity) close to the BH, it is necessary to extend
the standard (\( e=1 \)) tidal interaction formalism of Press \& Teukolsky
(\cite{Pre77}) to hyperbolic orbits. This is described in Appendix~\ref{sec:Tl}.

\begin{figure}[h]
{\centering \begin{tabular}{c}
\resizebox*{!}{0.4\textheight}{\includegraphics{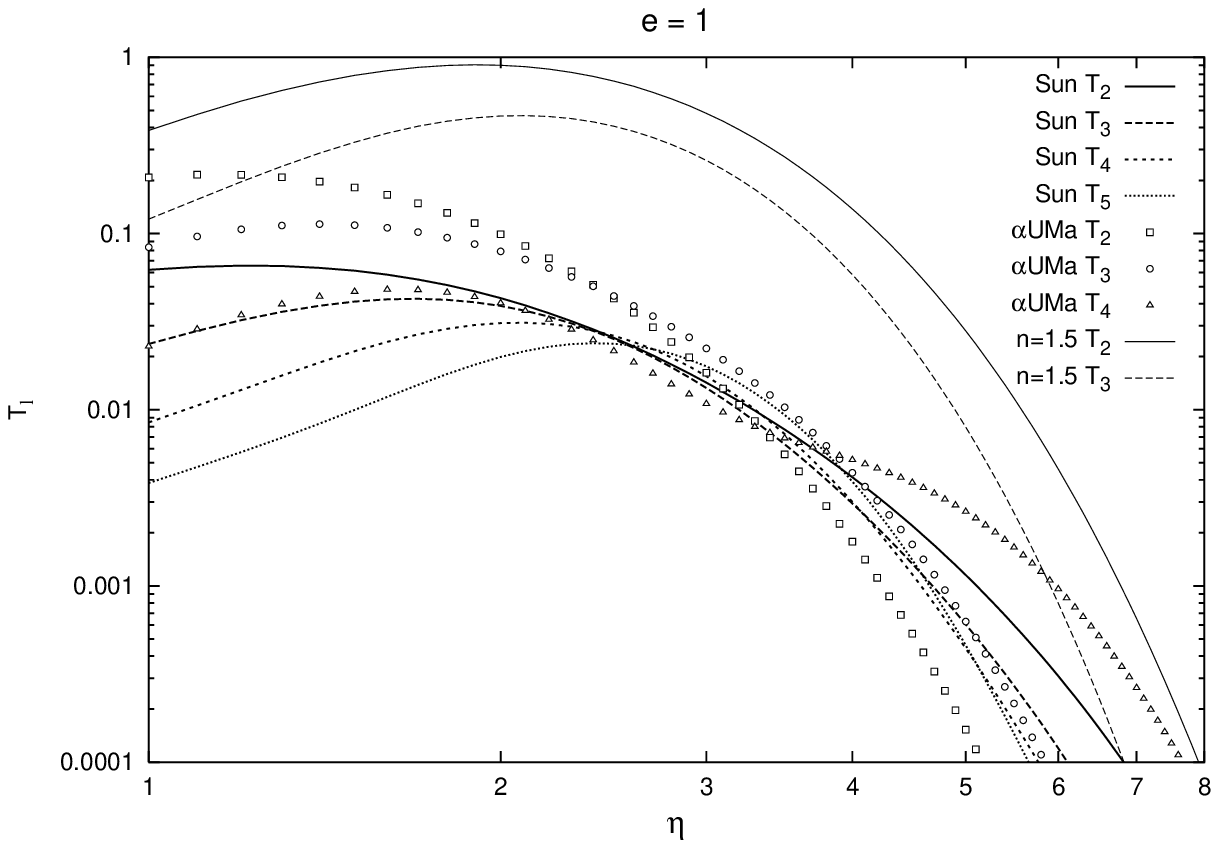}} \\
\resizebox*{!}{0.4\textheight}{\includegraphics{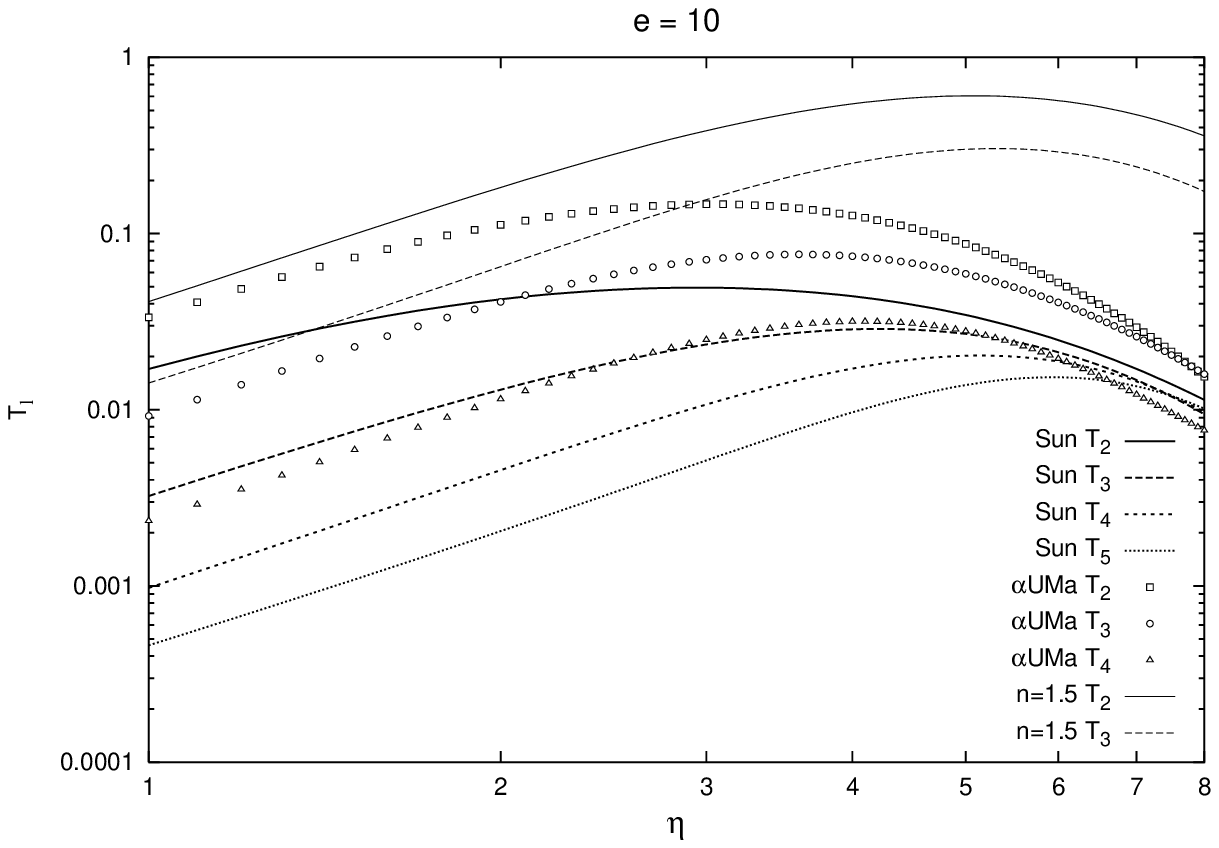}} \\
\end{tabular}\par}

\caption{\label{fig:Tl}The first few orders of \protect\( T_{l}\protect \) for the
Solar stellar structure model, the giant model and the ideal gas \protect\( n=1.5\protect \)
polytrope. Top: parabolic (\protect\( e=1\protect \)) encounters. Bottom: hyperbolic
(\protect\( e=10\protect \)) encounters.}
\end{figure}

In this work we consider two types of target stars: A MS dwarf and a red giant,
whose properties are summarized in table~\ref{tab: stype}. The detailed stellar
structures that we use to calculate the tidal coupling coefficients for these
stars are based on the solar model of Christensen-Dalsgaard et al. (\cite{Chr96})
and a model for the red giant \( \alpha  \)UMa (Guenther et al. \cite{Gue00}).~\texttt{}Figure~\ref{fig:Tl}
shows the run of \( T_{l} \) with \( \eta  \) for the two stars and for an
ideal gas \( n=1.5 \) polytrope, for parabolic (\( e=1 \)) and hyperbolic
(\( e=10 \)) orbits. We find that the tidal coupling in an ideal gas polytrope
is stronger than in either of the more realistic models. We also find a general
trend for the \( T_{l} \) coefficients to reach their maxima at larger values
of \( \eta  \) (larger \( \rp  \) for a given \( \mi  \)) with increasing
\( e \) (and increasing periastron velocity \( \vp  \)). This reflects the
fact that the coupling is strongest when \( \Wp =\vp /\rp  \) equals the lowest
frequency stellar pulsation mode.

\section{Spin-up by strong hyperbolic collisions}

\label{sec: strong}

\subsection{SPH simulations}

In order to extend the linear treatment of the soft encounters to strong (close
and penetrating) encounters, we simulated such encounters with the Smoothed
Particle Hydrodynamics (SPH) technique (Lucy \cite{Luc77}; Gingold \& Monaghan
\cite{Gin77}). In view of the difficulty of simulating all the aspects of a
real stellar collision, and in view of the many uncertainties in the details
of the stellar structure, the purpose of these simulations is not to calculate
\( \dW  \) precisely for specific collisions, but rather to gain \emph{qualitative}
insight about angular momentum transfer in strong encounters, which can then
be incorporated in our semi-analytic calculations by simple approximations.
The SPH code we use calculates the gravitational force by straight \( N^{2} \)
operations, and is therefore limited to relatively low resolution simulations
(typically \( N=2048 \) particles). The code integrates in time the entropy
equation (Hernquist \cite{Her93}); conserves particle momenta identically;
uses the artificial viscosity prescription given by Hernquist \& Katz (\cite{Her89},
Eqs. 2.22, 2.23, 2.37) and the time-step criteria of Katz, Weinberg \& Hernquist
(\cite{Kat96}). The amount of stellar mass-loss in the collisions is estimated
by the enthalpy criterion of Rasio \& Shapiro (\cite{Ras91}).

We verified the code by constructing stable \( n=1.5 \) polytrope configurations;
by confirming that the results converge as the number of particles is increased;
and by reproducing qualitatively the spin-up and mass-loss obtained in the SPH
simulations of Davies, Benz \& Hills (\cite{Dav91}), who used a much more realistic
stellar structure model. Of direct relevance is the fact that our SPH code reproduces
the results of the linear theory at the soft collision limit (\( \widetilde{r}_{p}\gtrsim 2.5 \))
and at smaller \( \rp  \) follows closely the SPH results obtained by Rasio
\& Shapiro (\cite{Ras91}) (Fig.~\ref{fig:dE}) in SPH simulations with \( 10^{4} \)
particles.

\subsection{Beyond the linear regime}

\subsubsection{Deep inelastic collisions}

\begin{figure}[h]
{\par\centering \resizebox*{!}{0.5\textheight}{\includegraphics{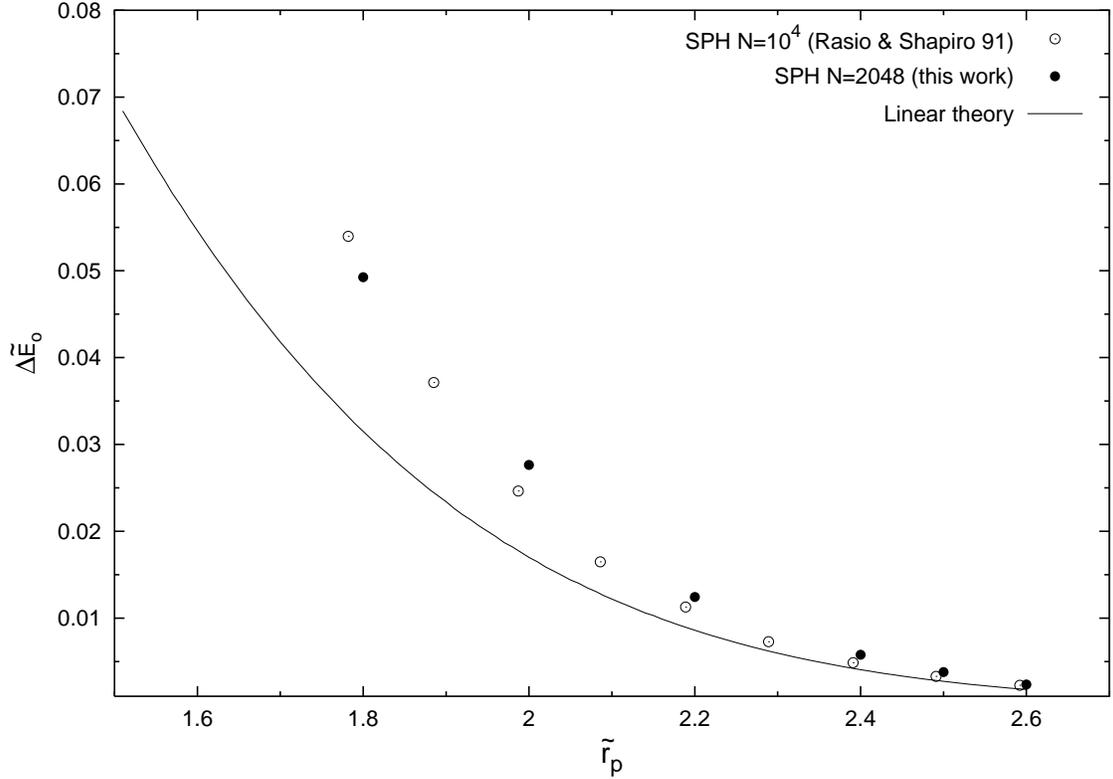}} \par}

\caption{\label{fig:dE}The energy transfered from the orbit to the star in a parabolic
collision between a point mass and an ideal gas \protect\( n=1.5\protect \)
polytrope of equal mass (\protect\( \mi =1\protect \), \protect\( \Eo =0\protect \))
as function of the peri-separation \protect\( \rp \protect \). The SPH results
from this work are compared with the higher resolution SPH results of Rasio
\& Shapiro (\cite{Ras91}) and with the predictions of the linear theory (Eq.~\ref{eq:dwtc}
with \protect\( C_{NL}=1\protect \)).}
\end{figure}

At small peri-separations (small \( \eta  \) for fixed \( \mi  \)), the sum
in Eq.~\ref{eq:dwt} converges slowly and the truncation of the \( T_{l} \)
series at some order \( l=k \) could under-estimate \( \dW  \). We make use
of the fact that the ratio \( T_{l+1}(\eta )/T_{l}(\eta ) \) is roughly constant
(Fig.~\ref{fig:Tl}) over the small-\( \eta  \) range of interest to extrapolate
the sum to high \( l \) by a geometric series
\begin{equation}
\label{eq:dwtc}
\dW \approx C_{NL}\frac{\widetilde{m}^{2}}{\MOI \Wp \rp ^{2}}\left( \sum ^{k-2}_{l=2}\frac{T_{l}}{\rp ^{2l}}+\frac{T_{k-1}}{\rp ^{2k-2}}\left[ 1-\frac{T_{k}}{\rp ^{2}T_{k-1}}\right] ^{-1}\right) \, ,
\end{equation}
where the constant \( C_{NL} \) is a non-linear correction factor, which we
calibrate by the SPH simulations as discussed below. Figure~\ref{fig:dE} compares
the prediction of the linear theory with the numeric SPH results without correcting
for non-linear effects (\( C_{NL}=1 \)). While the two agree at the limit of
soft (distant) encounters, \( \dE  \) grows faster than the linear theory at
close encounters. In this particular example the high-order correction is very
small, and does not exceed \( 2\% \) down to \( \rp =1.6 \).

\begin{figure}[h]
{\par\centering \resizebox*{!}{0.6\textheight}{\rotatebox{270}{\includegraphics{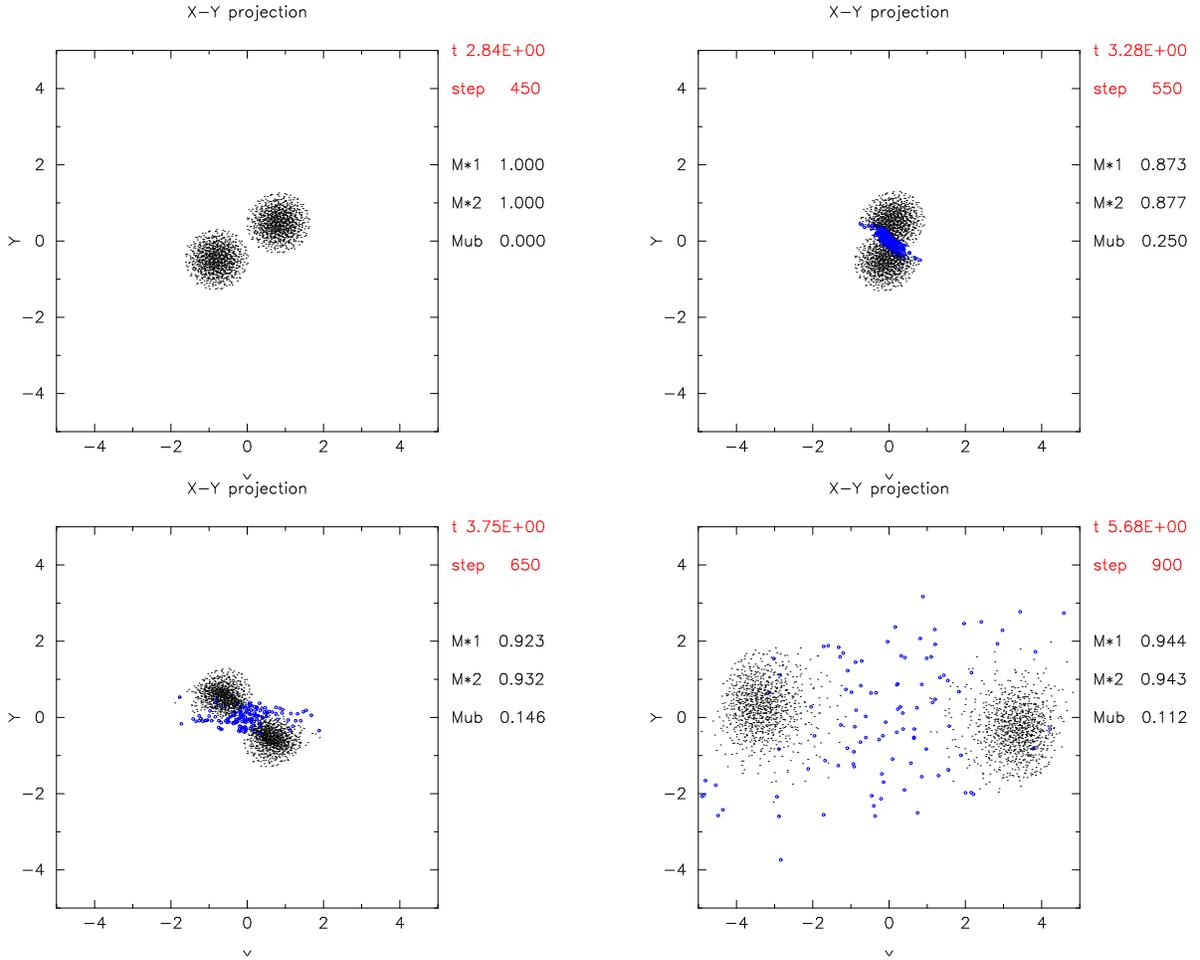}}} \par}

\caption{\label{fig:2poly}A sequence of snapshots from an SPH simulation of a deep
collision between two equal mass stars (modeled as ideal gas \protect\( n=1.5\protect \)
polytropes with \protect\( N=1024\protect \) particles each). The orbital parameters
of the encounter are \protect\( \rp =1.0\protect \) and \protect\( \Eo =2\protect \).
At \protect\( \widetilde{t}=20\protect \) after the periastron passage (not
shown here), the stars have lost 5\% of their mass each, were spun-up by \protect\( \dW =0.025\protect \)
and acquired a moment of inertia \protect\( \MOI =0.47\protect \), more than
twice the initial value.}
\end{figure}

\begin{figure}[h]
{\par\centering \resizebox*{!}{0.6\textheight}{\includegraphics{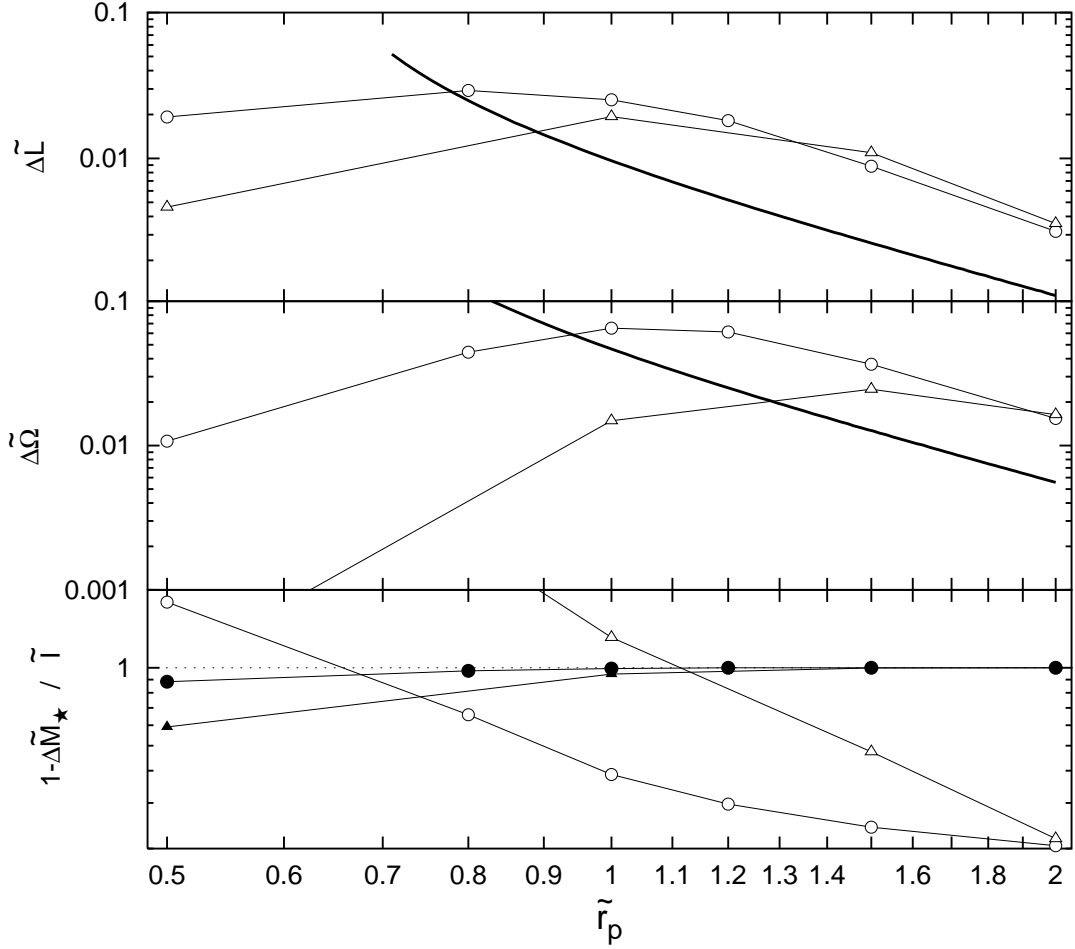}} \par}

\caption{\label{fig:deep}The response of a star (\protect\( \mi =1\protect \), modeled
as an ideal gas \protect\( n=1.5\protect \) polytrope) to deep non-linear hyperbolic
collisions (\protect\( \Eo =2\protect \)) with another star, for both the case
where the impactor (modeled as a point mass) is compact relative to the target
star (circles) and the case where the impactor (modeled as an ideal gas \protect\( n=1.5\protect \)
polytrope) is of the same size as the target star (triangles). The SPH results
(thin lines with symbols) are compared with the analytic calculations without
the non-linear correction (thick lines) for \protect\( \dL \protect \) (top
panel) and \protect\( \dW \protect \) (middle panel). Also shown in the bottom
panel are \protect\( 1-\Delta \widetilde{M}_{\star }\protect \), the mass that
remains bound to the star after the collision (thin lines with filled symbols)
and \protect\( \MOI \protect \), the star's moment of inertia after the collision
(thin lines with open symbols).}
\end{figure}

\begin{figure}[h]
{\par\centering \resizebox*{!}{0.6\textheight}{\rotatebox{270}{\includegraphics{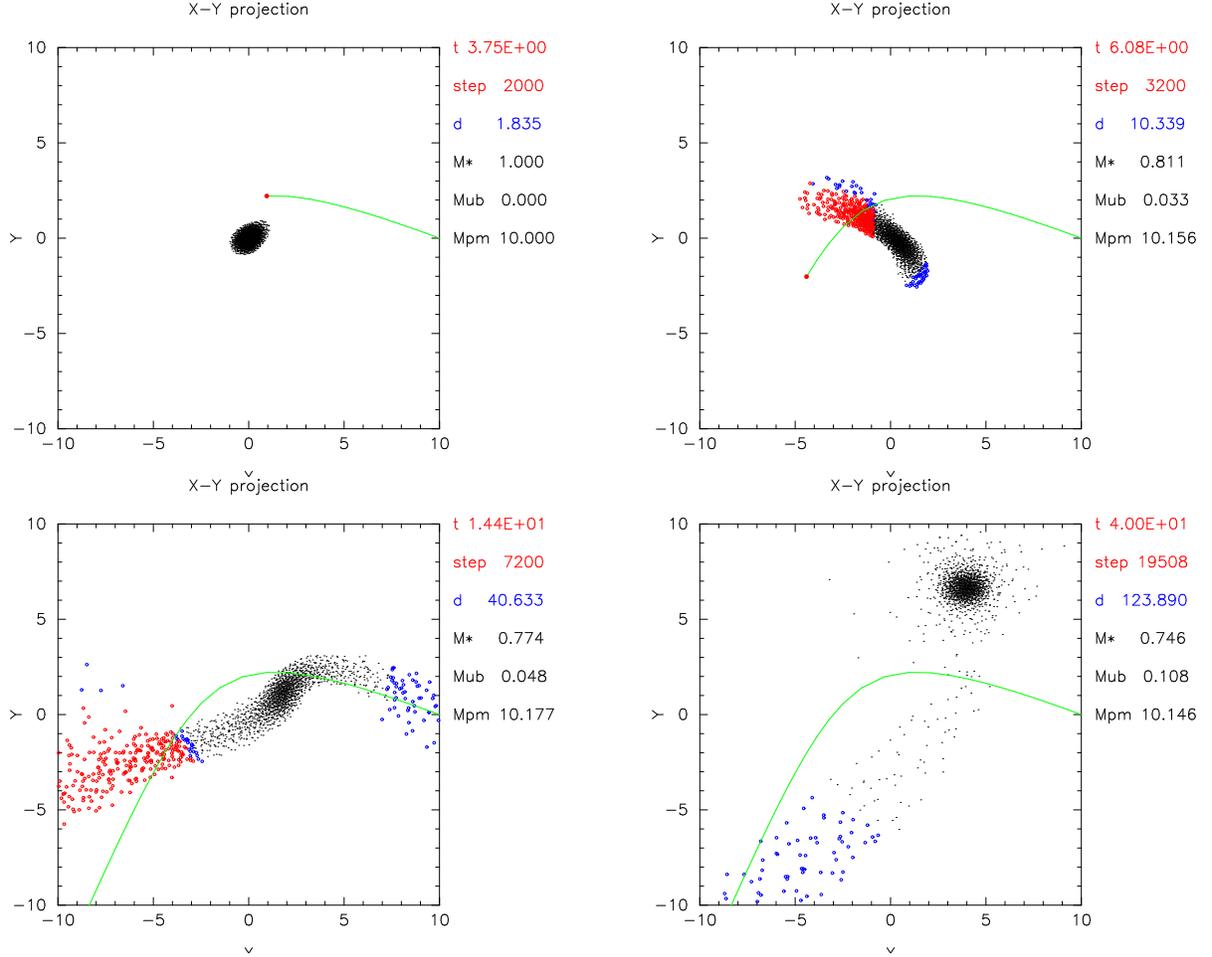}}} \par}

\caption{\label{fig:rt}A sequence of snapshots from an SPH simulation of a collision
at the tidal limit (Eq.~\ref{eq:rt}) between a \protect\( \mi =10\protect \)
BH (modeled as a point mass) and a MS star (modeled as an ideal gas \protect\( n=1.5\protect \)
polytrope). The dots represent SPH gas particles that remain bound to the star,
and the circles represent those that are lost. The line traces the trajectory
of the BH in a frame where the target was initially at rest. The orbital parameters
of the encounter are \protect\( \rp =\rt =2.0\protect \) and \protect\( \Eo =1.0\protect \).
In spite of the significant tidal stretching of the star (bottom left panel),
the star ultimately relaxes after losing 25\% of its mass. At \protect\( \widetilde{t}=124\protect \)
(bottom right panel) the stellar mass that has settled back to within \protect\( \widetilde{r}=2\protect \)
of the stellar core rotates at 20\% of the centrifugal break-up velocity. The
moment of inertia of the mass within \protect\( \widetilde{r}=2\protect \)
is \protect\( \MOI =0.45\protect \), more than twice its initial value.}
\end{figure}

Figures.~\ref{fig:2poly} and \ref{fig:rt} show snapshots from two SPH simulations
of extremely non-linear collisions. Figure~\ref{fig:deep} shows the run of
the dynamical properties of the target star with \( \rp  \) in \( \mi =1 \),
\( \Eo =2 \) collisions. Our analytic expressions for \( \dL  \) and \( \dW  \)
consistently underestimate the SPH results for a point mass impactor by a factor
of \( \sim 3 \) down to \( \rp \sim 1.5 \). Similarly, it under-estimates
\( \dE  \) in parabolic collisions by a factor \( \sim 1.5 \) at \( \rp =1.8 \)
(Fig.~\ref{fig:dE}). We find that at smaller peri-separations where significant
amounts of mass are ejected in the collisions, a large fraction of the angular
momentum that is taken out of the orbit is deposited in the ejecta rather than
in the target star. This truncates the \( \dW \sim \rp ^{-5} \) divergence
and limits the spin-up efficiency, so that the difference between the analytic
calculations and simulation results decreases with \( \rp  \) until the two
cross over. We approximate the decrease in spin-up efficiency by introducing
a truncation peri-separation \( \widetilde{r}_{0} \) such that \( \dW (\widetilde{r}_{p})=\dW (\widetilde{r}_{0}) \)
for \( \rp <\widetilde{r}_{0} \). 

We now discuss our choice of \( \widetilde{r}_{0} \). Figure~\ref{fig:deep}
shows that \( \dW  \) is much more suppressed than \( \dL  \) at small \( \rp  \)
because the stellar moment of inertia increases as a consequence of the collision.
The increase in \( \MOI  \) far exceeds the small change that is expected due
to the stellar oblateness that develops in response to the rotation. The effect
seen in the simulation is due to the heating and subsequent expansion of the
star by the collision. Our SPH models do not include the radiative processes
that are necessary for describing the later stages of cooling and contraction.
It is therefore likely that the final value of \( \dW  \) is not as strongly
suppressed as is implied by Figure~\ref{fig:deep}, but rather follows more
closely the behavior of \( \dL  \). A collision with an impactor of a size
comparable to that of the target results in much more mass-loss than a collision
with an impactor that is effectively a point mass (e.g. a stellar remnant on
a MS star or a MS star on a giant). However, it is likely that the SPH results
over-estimate the mass-loss since a \( n=1.5 \) polytrope has a significantly
lower binding energy than the realistic stellar structure models on which we
base our analytic calculations (Table~\ref{tab: stype}). Figure~\ref{fig:deep}
shows that the analytic estimate of \( \dW  \) equals the value derived from
the simulation at \( \rp \sim 1.3 \) and \( \rp \sim 0.9 \) for polytrope
and point mass impactors, respectively. For \( \dL  \) the analytic estimate
and the simulation results are equal at \( \rp \sim 0.9 \) and \( \rp \sim 0.8 \),
respectively. Based on the arguments presented above, we adopt in this work
the simple prescription that \( \widetilde{r}_{0}=1.0 \) or the size of the
impactor, whichever is larger, for all types of collisions. We note that the
analytic calculations still \emph{under-estimate} the SPH results by a factor
of 2 at \( r_{0} \), and by a factor of 3 at \( \rp =2 \). A difference of
a similar magnitude is seen also in parabolic collisions (Fig.~\ref{fig:dE}).
We compensate for this discrepancy by setting the non-linear correction factor
to \( C_{NL}=2 \). A more extensive investigation of parameter space by SPH
simulations with more realistic stellar structure models will be required to
refine this prescription.

\subsubsection{Prompt disruption of tidally formed binaries}

Close hyperbolic encounters of stars in the low energy tail of the orbital energy
distribution can lead to the formation of a bound system (\( \Eo +\Delta \Eo <0 \))
with a large semi-major axis \( \widetilde{a} \). Although such encounters
are very efficient in spinning-up stars, the subsequent evolution of the stellar
rotation in binaries is very different from that due to stochastic encounters.
We do not consider such cases in this work and they are not included in the
average \( \dWm  \) (Eq.~\ref{eq:edwm}). Because of the very large stellar
density and the proximity of the massive BH, it is necessary to check whether
a newly formed binary can survive its first orbital period without being disrupted
by a tidal interaction with either a third nearby star or the central BH. 

We take this into account by considering the orbit as bound only if the change
in the orbital angular momentum, \( \Delta \widetilde{L}_{\mathrm{orb}} \),
due to the torque exerted on the system by the third mass \( \widetilde{m}_{3} \)
(star or central BH) is smaller than the orbital angular momentum \( \widetilde{L}_{\mathrm{orb}} \),

\begin{equation}
\label{eq:Lorb}
\dL _{\mathrm{orb}}=4\pi \left( \frac{\widetilde{a}}{\widetilde{d}}\right) ^{3}\left( \frac{\mi _{3}}{1+\mi }\right) \left( 1-e^{2}\right) ^{-1/2}\widetilde{L}_{\mathrm{orb}}\, ,
\end{equation}
where \( \widetilde{d} \) is the distance of \( \mi _{3} \) from the binary
and where we used the relation \( \widetilde{L}^{2}_{\mathrm{orb}}=\mi \mr \widetilde{a}(1-e^{2}) \).
For disruption by a star, \( d\sim \ns ^{-1/3} \), \( \mi _{3}\sim \mi  \)
and the no-disruption criterion is 
\begin{equation}
4\pi \ns \widetilde{a}^{3}\mr \left( 1-e^{2}\right) ^{-1/2}<1\, .
\end{equation}
The eccentricity and the semi-major axis are estimated from the unperturbed
peri-separation and the bound orbit's energy
\begin{equation}
e=\frac{2(\Eo +\Delta \Eo )}{\mr }\rp +1\, ,\qquad \widetilde{a}=\frac{\rp }{1-e}\, .
\end{equation}
In practice, we find that the contribution from disrupted binaries to \( \delta \widetilde{\Omega } \)
in the GC is negligible.

\subsubsection{Tidal and collisional destruction}

The precise criterion for tidal destruction of stars by hyperbolic encounters
is not well known. In this work we adopt the simple criterion that tidal break-up
occurs when \( \Delta \widetilde{v} \), the change in the velocity of a mass
element on the stellar surface, exceeds the escape velocity, \( \Delta \widetilde{v}\sim 2\mi /\rp ^{3}\times \rp /\vp >\sqrt{2} \).
The tidal radius \( \rt  \) is a function of the impactor mass and orbital
energy and is given by the solution to the equation 
\begin{equation}
\label{eq:rt}
\rt =\left( \frac{\mr \mi }{\mi +\rt \Eo }\right) ^{1/3}\mi ^{1/3}\, .
\end{equation}
 Our SPH simulations indicate that this is a conservative criterion. Figure
\ref{fig:rt} shows a sequence of snapshots from an encounter with \( \rp =\rt =2.0 \),
\( \Eo =1.0 \) between a \( \mi =10 \) black hole (modeled as a point mass)
and a MS dwarf (modeled as an ideal gas \( n=1.5 \) polytrope). The star survives
the collision after suffering \( \sim 25\% \) mass-loss. We performed several
spot-checks with different values of \( \rp  \) and \( \Eo  \) and verified
that this prescription for \( \rt  \) indeed roughly demarcates the boundary
where the fractional mass-loss increases to order unity, and that \( \dW  \)
significantly exceeds the value predicted by our extrapolated linear formalism
for such deep collisions. 

When \( \rt <1 \) (point mass impactor) or \( \rt <2 \) (impactor of same
size as target) it is necessary to consider the possibility of collisional destruction.
Our SPH simulations indicate that the \( n=1.5 \) polytrope stellar models
can survive very deep collisions (\( \rp \sim 0.5 \), cf Figs~\ref{fig:2poly},
\ref{fig:deep}), albeit with a significant mass-loss. One concern when considering
penetrating collisions by compact remnants is the energy release by nuclear
reactions near the surface of the impactor. R\'{o}\.{z}yczka et al. (\cite{Roz89})
and Ruffert \& M\"{u}ller (\cite{Ruf90}) find that nuclear reactions probably
do not play a significant role even in parabolic head-on collisions between
low-mass stars and white dwarfs. We will assume that stars can survive collisions
down to \( \rp =0.5 \) even when the impactor is a compact object.

\subsubsection{The survival probability}

In an environment that is dense enough for efficient tidal spin-up there is
also a non-negligible probability for destructive head-on collisions. Our estimate
of \( \dW  \) implicitly assumed that the target star survives its full life
span \( \Tt  \). We now estimate the survival probability of stars against
a close collision with some peri-separation \( \rc  \) in order to check what
fraction of the star can survive long enough to acquire significant rotation.
For simplicity, we omit gravitational focusing, which is negligible for long-lived
MS stars very close to a massive BH (\( \si \gg 1 \)).

The collisions are a random process in both \( \rp  \) and in the time of the
periastron passage. The rate for collisions with peri-separation in the range
\( \rp  \) to \( \rp +d\rp  \) is 
\begin{equation}
d\qc =\ns \vi 2\pi \rp d\rp \, ,
\end{equation}
where in a Maxwellian DF with 1D velocity dispersion \( \si  \) the rms \( \vi  \)
is \( \sqrt{6}\si  \). The probability density function \( \rho _{c} \) (pdf)
for a star to have over a time interval \( \Tt _{c} \) a closest encounter
with peri-separation \( \rp  \) in the range \( \left[ \rc ,\rc +d\rc \right]  \)
is given by
\begin{equation}
\rho _{c}d\rc =\exp \left( -\pi \ns \vi \Tt _{c}\rc ^{2}\right) 2\pi \ns \vi \Tt \rc d\rc \, ,
\end{equation}
where the first term is the Poisson probability for avoiding an encounter with
\( \rp <\rc  \) and the second is the probability for having at least one in
the required range. The fraction of target stars, \( f_{c} \), that avoid a
collision with peri-separations \( \rp <\rc  \) (the survival probability)
can be written in terms of \( \left\langle \rc \right\rangle =\left( 4\ns \vi \Tt _{c}\right) ^{-1/2} \)
as 
\begin{equation}
\label{eq:fc}
f_{c}=\int _{\widetilde{r}_{c}}^{\infty }\rho _{c}d\rp =\exp \left( -\frac{\pi }{4}\left[ \frac{\rc }{\left\langle \rc \right\rangle }\right] ^{2}\right) \, .
\end{equation}

\section{Tidal spin-up in the inner Galactic Center}

\label{sec: results}

Up to this point our treatment of the tidal spin-up effect was general. We now
turn our attention to the specific case of the GC.

\subsection{The stellar velocity distribution in the GC}

\label{sec:vDF}

The effects of the stellar collisions depend critically on the relative velocity
of the two colliding stars and their mass ratio. It is therefore important to
understand the mass dependence of the velocity distribution. We are interested
in particular in the case where the stellar system around the black hole has
undergone two-body relaxation, as appears to be the situation in the GC (Alexander
\cite{Ale99a}). 

Let \( f_{m}(\epsilon ) \) be the DF of stars of mass \( m \) (\( m_{1}\le m\le m_{2} \))
as function of specific energy in a spherical stellar system whose potential
is dominated by a central BH, where \( \epsilon =\Psi -v^{2}/2>0 \), \( \Psi =GM_{\bullet }/r \),
and \( M_{\bullet } \) is the mass of the BH. Bahcall and Wolf (\cite{Bah77})
have shown that when such a system undergoes two-body relaxation, the DF has
the following properties: \( p_{m}\equiv d\ln f_{m}/d\ln \epsilon \sim \mathrm{const}. \)
to a good approximation; \( p_{m_{2}}\simeq 1/4 \); and \( p_{m}/p_{m_{2}}\simeq m/m_{2} \)
so that \( 0\lesssim p_{m}\lesssim 1/4 \). In this case \( f_{m}(\epsilon )\propto \epsilon ^{p_{m}} \)
and the velocity dispersion \( \sigma ^{2}_{m}=\left\langle v^{2}\right\rangle /3 \)
is

\begin{equation}
\label{eq:sigm}
\sigma _{m}^{2}=\left( \frac{1}{p_{m}+5/2}\right) \frac{GM_{\bullet }}{r}\, .
\end{equation}
 The velocity dispersion is almost independent of the stellar mass. The relative
change in the value of \( \sigma _{m}^{2} \) over the full mass range is only
\( \sim 10\% \) independently of the ratio \( m_{2}/m_{1} \), in marked contrast
with the wide spread of velocities expected in the case of equipartition where
\( \sigma _{m}^{2}\propto m^{-1} \). The reason why the relaxed system does
not reach equipartition can be understood by considering the fate of a massive
star that is momentarily on a circular orbit. The orbital radius depends only
on the specific energy. Equipartition works to equate the kinetic energy per
star, thereby always reducing the specific energy of the massive stars and causing
them to sink to ever lower orbits. This is analogous to the equipartition instability
discussed by Spitzer (\cite{Spi69}) in the context of a stellar cluster without
a central BH, but where the fraction of mass in the massive stars is large enough
to create a centrally concentrated sub-system.

\begin{figure}[h]
{\par\centering \resizebox*{!}{0.5\textheight}{\includegraphics{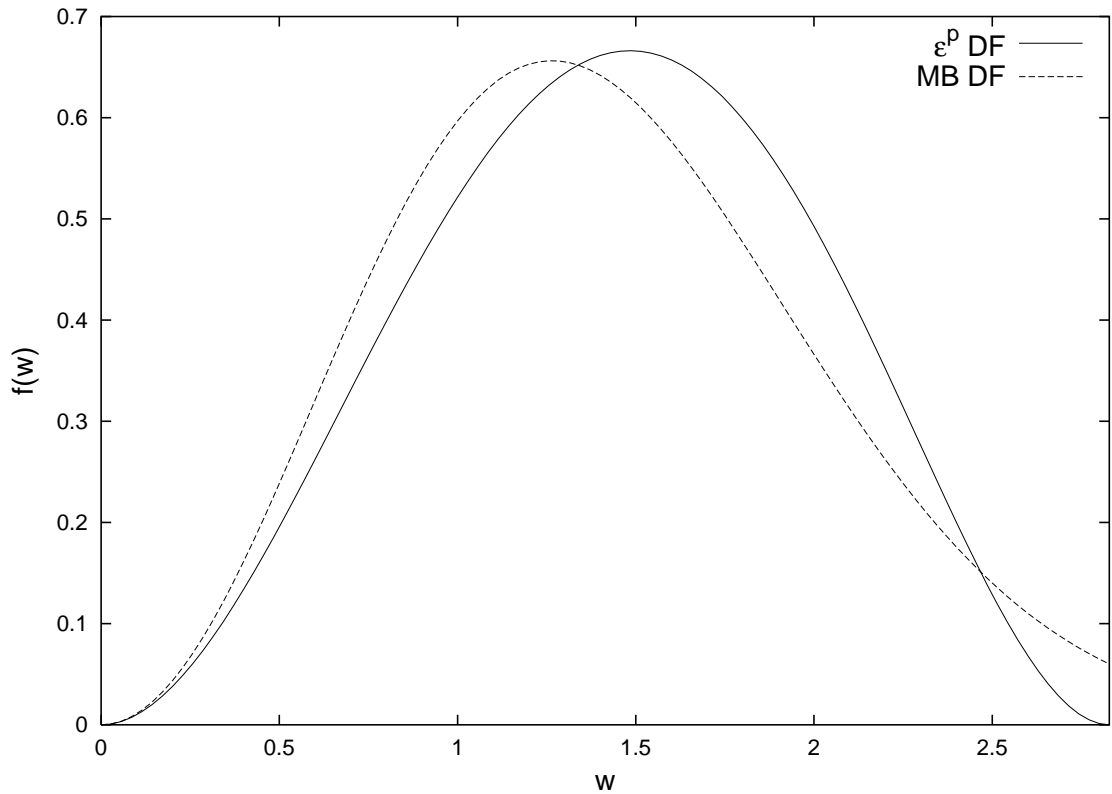}} \par}

\caption{\label{fig:v2DF}A comparison of the DFs of the relative stellar velocity (\protect\( w\equiv v_{2}/\Psi ^{1/2}\protect \))
in a Maxwellian DF and of a \protect\( \epsilon ^{p}\protect \) DF of relaxed
low-mass stars very near a black hole (Bahcall \& Wolf \cite{Bah77}). Both
DFs have the same 1D velocity dispersion and the same mean orbital energy.}
\end{figure}

The resulting distribution of the relative velocity \( \bmath {v_{2}}=\bmath {v_{a}}-\bmath {v_{b}} \)
between two stars of masses \( m_{1}\le m_{a}\le m_{b}\le m_{2} \) is given
by 
\begin{equation}
\label{eq:DF2}
f(\bmath {v_{2}})=\frac{1}{8\pi ^{3}\Psi ^{3/2}}\frac{\Gamma (\frac{5}{2}+p_{m_{a}})\Gamma (\frac{5}{2}+p_{m_{b}})}{\Gamma (1+p_{m_{a}})\Gamma (1+p_{m_{b}})}\int d^{3}u\left( 1-\frac{u^{2}}{2}\right) ^{p_{m_{a}}}\left( 1-\frac{\left( \bmath {u}-\bmath {w}\right) ^{2}}{2}\right) ^{p_{m_{b}}}\, ,
\end{equation}
where \( \bmath {u}\equiv \bmath {v_{a}}/\Psi ^{1/2} \) and \( \bmath {w}\equiv \bmath {v_{2}}/\Psi ^{1/2} \)
and the integration is over the region \( u<\sqrt{2} \) and \( \left| \bmath {w}-\bmath {u}\right| <\sqrt{2} \).
This distribution is not very different from a Maxwellian DF with the same 1D
velocity dispersion \( \sigma  \). This can be seen in Fig.~\ref{fig:v2DF},
which compares the Maxwellian DF with the case where both \( p_{m_{a}}=p_{m_{b}}=0 \)
(low mass stars) and Eq.~\ref{eq:DF2} simplifies to
\[
f(v_{2})dv_{2}=\frac{\Gamma (5/2)^{2}}{24\pi }w^{2}\left( 32\sqrt{2}-24w+w^{3}\right) dw\, .\]
In particular, the two DFs have the same mean orbital energy 
\begin{equation}
\label{eq:avEo}
\left\langle E_{o}\right\rangle =3\mu \sigma ^{2}\, .
\end{equation}

We conclude that the velocity field in a relaxed stellar system very near a
BH is well approximated by a Maxwellian velocity distribution where the 1D velocity
dispersion is independent of the stellar mass.

\subsection{The stellar population in the GC }

Stellar population synthesis models of the observed luminosity function averaged
over the inner few parsecs of the GC (Alexander \& Sternberg \cite{Ale99b})
indicate that it is well described by a continuous star forming population with
a Miller-Scalo initial mass function (IMF) (Miller \& Scalo \cite{Mil79}) with
masses in the range \( 0.1 \)--\( 125\, M_{\odot } \). The mean impactor mass
(live stars and remnants) in this model is \( \left\langle m\right\rangle =0.5\, M_{\odot } \).
In an isolated system the present day mass function (PMF) preserves the IMF
distribution for low mass stars that are longer-lived than the system, but falls
more rapidly for the shorter-lived massive stars, so that over time the low-mass
stars accumulate and take an ever larger fraction of the total stellar mass.
However, the timescale for mass segregation in the GC, which is of the same
order as the relaxation timescale, is only \( 3\, \mathrm{Gyr} \) (e.g. Alexander
\cite{Ale99a}). It is therefore reasonable to assume that the mass fraction
of the very low-mass stars in the inner GC is significantly lower than implied
by continuous star formation.

In view of the uncertainties in the low-mass end of the PMF, we do not attempt
to construct a detailed mass function for the impactors in the innermost GC.
Instead, we base it on the stellar synthesis model (table~\ref{tab:rem}) and
take account of the mass-segregation by assuming a Salpeter power-law PMF with
a low-mass cutoff close to the mean impactor mass. The model parameters are
listed in table~\ref{tab:PMF}. The mean impactor mass in this model is \( \left\langle m\right\rangle =0.9\, M_{\odot } \).
We carried out spot-checks to verify that the exact values of the mass ranges
and power-law indices do not affect the final results significantly. The one
important assumption in this model is that the mass in the inner GC is not dominated
by very low-mass stars, which are inefficient in raising tides. More detailed
modeling of the dynamical evolution of the inner GC will be required to verify
this. The results for the tidal spin-up that are presented below can easily
be scaled to other mass fraction ratios through Eqs.~\ref{eq:dq} and \ref{eq:edwm}.

We represent the mass distribution in the innermost GC by a stellar cusp of
the form (Sec.~\ref{sec:vDF} here; Alexander \cite{Ale99a}) 

\begin{equation}
\rho =10^{6}\left( \frac{r}{0.4\, \mathrm{pc}}\right) ^{-(3/2+p_{m})}\, M_{\odot }\, \mathrm{pc}^{-3}\, ,
\end{equation}
and the 1D velocity dispersion by Eq.~\ref{eq:sigm}. We set \( p_{m}=0 \)
to represent the typical low-mass impactors and assume \( M_{\bullet }=2.6\times 10^{6}\, M_{\odot } \)
(Genzel et al. \cite{Gen97}).

\subsection{Results}

We calculated the spin-up in the GC following the procedure outlined in the
previous sections. To summarize, the calculation proceeded by a triple numeric
integration. First, integration over the impactor mass function (table~\ref{tab:PMF}).
Second, integration over \( \rp  \) from a suitably large distance down to
the larger of \( \rt  \) (Eq.~\ref{eq:rt}) and \( \rp =0.5 \) where for \( \rp <\widetilde{r}_{0}=1 \),
\( \dW  \) was held fixed to account for mass-loss. Third, integration over
\( \Eo  \) from a suitable large value down to 0. Collisions that resulted
in bound orbits (Eq.~\ref{eq:Lorb}) were omitted from the sum. For each point
in the integration, \( \dW  \) was calculated (Eq.~\ref{eq:dwtc}) with a non-linear
correction factor \( C_{NL}=2 \) and summed in quadrature, weighted by the
differential rate (Eq.~\ref{eq:dq}). We assumed that long after the collision
\( \MOI  \) recovers its initial value and that \( \mi  \) is unchanged. 

\begin{figure}[h]
{\par\centering \resizebox*{!}{0.5\textheight}{\includegraphics{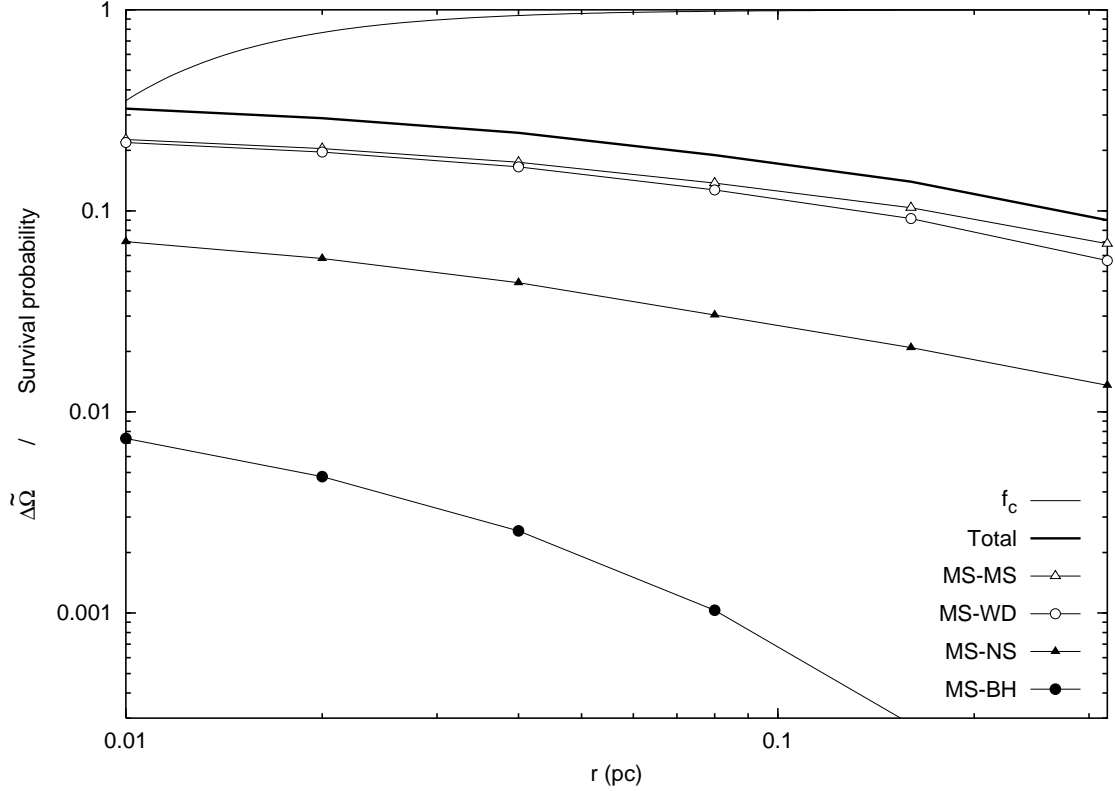}} \par}

\caption{\label{fig: dwgc}The mean stellar spin \protect\( \dW \protect \) of a MS
star as function of distance from the BH in the GC assuming a Salpeter PMF and
the remnant mass fractions listed in table~\ref{tab:rem}. The survival probability
against a destructive collision with \protect\( \rp <0.5\protect \) over 10
Gyr (Eq.~\ref{eq:fc}) is also plotted for a mean impactor mass of \protect\( \left\langle m\right\rangle =0.9\, M_{\odot }\protect \). }
\end{figure}

Figure~\ref{fig: dwgc} shows the run of \( \dWm  \) with distance from the
BH in the GC for the model MS target star with \( T=10 \) Gyr, as well as the
separate contributions from collisions with white dwarfs (WDs), neutron stars
(NSs) and stellar mass BHs. Figure~\ref{fig: dwgc} also shows that the survival
probability is almost unity as close as 0.02 pc to the BH, and so the fact that
it was not taken into account explicitly in the estimate of \( \dWm  \) does
not introduce a serious error. The stellar rotation falls only slowly with distance
from the black hole and is at the level of \( \dWm \sim 0.1 \)--\( 0.3 \)
in the inner 0.3 pc. Most of the effect comes from collisions with MS star and
WDs. Field MS stars later than \( \sim \! \mathrm{F}5 \) are very slow rotators
(Gray \cite{Gra92}). For \( \sim \! 1\, M_{\odot } \) stars these values of
\( \dWm  \) correspond to rotational velocities 20--60 times higher than normal
(Table~\ref{tab: stype}).

We carried out similar calculations for giant stars. We find that the spin-up
is much smaller, \( \dWm \sim 0.01 \)--\( 0.02 \), because of the much higher
\( \si  \) and the shorter lifetime in the giant phase. This is similar to
the rotational velocity of field giants later than \( \sim \! \mathrm{G}3 \),
and thus tidal spin-up may increase their rotation to double the normal value.

\section{Discussion and summary}

\label{sec: discuss}

Dense stellar cusps around massive BHs are environments where stellar collisions
are frequent and energetic. Hyperbolic head-on collisions may destroy the colliding
stars, but are rare. For every head-on collision there are many more close tidal
encounters and grazing collisions. The cumulative effects of such encounters
are more subtle. Energy and angular momentum are transfered from the orbit to
the stars, during the fly-by the stellar structure may be significantly disturbed
by the tidal forces and some mass may be lost. However, the stars survive the
collision, and because of the high initial orbital energy, they rarely form
a bound binary system. This is in contrast to the situation in dense globular
clusters. The stellar dynamical and thermal relaxation timescales are very short
relative to the stellar lifetime and the star can radiate the excess energy
quickly. It is harder to shed the excess angular momentum, since magnetic breaking
operates on timescales of order of the stellar lifetime. Thus the direct long
term effects of the collision are increased rotation, and possibly some mass-loss
and some mixing of the stellar envelope.

In this study we calculated the magnitude of the rotation that is built up in
a random walk fashion as the star undergoes multiple hyperbolic tidal encounters.
This is of interest because of the effects high rotation may have on the star's
evolution and on its observed properties and because, unlike destructive collisions,
the spin-up affects the entire stellar population and extends over a much larger
volume of the galactic nucleus. Many, if not most galaxies have a super-massive
BH in their nucleus, and so high stellar rotation in galactic nuclei may be
common.

Our approach to the problem was to use detailed stellar structure models to
calculate, for the first time, the tidal coupling constants for arbitrary hyperbolic
orbits in the linear regime of soft encounters. Because the tidal energy and
angular momentum fall off as a high power of the peri-separation, it was necessary
to extend the calculation to the strongly non-linear regime of grazing and penetrating
collisions. We carried out a suite of SPH simulations to study the qualitative
behavior of such collisions and then conservatively extrapolated the exact linear
calculations to the non-linear regime by several simple prescriptions. The simulations
indicated the following. (1) At peri-separations closer than about twice the
target star's radius, the linear results are smaller than the actual spin-up
by a factor of at least two. (2) Once mass-loss becomes significant, the ejecta
carries away a large fraction of the angular momentum that is extracted from
the orbit, and therefore the spin-up saturates at its level just before the
onset of mass-loss. (3) Stars can survive deep collisions down to a peri-separation
of half the target star's radius. (4) A simple tidal disruption criterion for
hyperbolic encounters can roughly indicate the point where the fractional mass-loss
increases to order unity.

We calculated the spin-up of stars in the inner parsec of the Galactic Center,
which is of special interest because of the high quality and wealth of details
of the observed stellar data. We find that over 10 Gyr, \( \sim \! 1\, M_{\odot } \)
MS stars in the inner 0.3 pc are stochastically spun-up to 10--30\% of the centrifugal
break-up velocity (20--60 times higher than is usual for such stars in the field).
This effect decreases only weakly with distance from the BH because the increased
tidal coupling at lower collision velocities largely compensates for the decrease
in the collision rate at lower stellar densities. We estimated also the stellar
survival probability against head-on collisions over 10 Gyr, and found that
it was significantly large even very close to the center. The fact that the
spin-up is roughly constant over the volume of the inner GC implies that no
large error was introduced by neglecting the fact that stars on non-circular
orbits sample a varying stellar density over their lifetime. The spin-up of
giant stars over the giant phase is much smaller, or order 1--2\% of the centrifugal
break-up velocity (doubling the rotation that is usual for such stars in the
field).The effect is smaller because of their short lifetime and because of
the large ratio between the collision velocity and the stellar escape velocity,
which decreases the tidal coupling.

These results suggest that stochastic spin-up is an important stellar effect
in BH cusps. However, several caveats apply. This results depend on the validity
of our extrapolation to the non-linear regime. A real star is not well represented
by an ideal gas \( n=1.5 \) polytrope, and more realistic and extensive hydrodynamical
simulations, including the effects of radiation and nuclear burning, will be
needed to verify our results. We note however that the \( n=1.5 \) polytrope
model is significantly less bound than the realistic stellar models (Table~\ref{tab: stype}).
This implies that our estimates for the minimal peri-separation and the tidal
radius may be overly conservative. In addition, because a smaller binding energy,
larger tidal coupling coefficients (Fig.~\ref{fig:Tl}) and a larger moment
of inertia go hand in hand, the spin-up (\( \dW \sim T_{l}/\MOI  \)) is less
sensitive to the details of the stellar structure than any of these quantities
separately. Another uncertainty in applying our results to any specific system,
such as the GC, lies in modeling the PMF of the impactors. Generally, mass segregation
will work towards increasing the spin-up effect by pushing the ineffective low
mass projectiles out of the central region. Dynamical models of the evolution
of the galactic nucleus are required to put this on a quantitative footing.
We considered the survival probability of the stars against head-on collisions,
but there are also other effects that may compete against the spin-up, which
were not included in our estimate. Magnetic breaking, although slow, may be
effective over 10 Gyr. Unfortunately, at present the details of this process
can not be modeled with any certainty. The cumulative mass-loss in the course
of many non-linear collisions was also not considered.

Detailed predictions for the observational consequences of high rotation are
outside the scope of this work. We limit our comments on this matter to noting
that rotation lowers the effective temperature and luminosity of a star (Kippenhahn,
Meyer-Hofmeister \& Thomas \cite{Kip70}), but does not significantly affect
the spectral classification of a star until it is close to break-up (Gray \cite{Gra92}).
The long term effects of rotation on stellar evolution may be more significant.
We calculated the effective angular velocity, assuming solid body rotation.
The actual distribution of angular momentum could become stratified over time
and lead to rotational support of the core (VandenBerg et al. \cite{Van98})
or to the replenishment the hydrogen in the core by large scale deep mixing
(Sweigart \cite{Swe97}). These effects will manifest themselves in the giant
phase of the stars. Finally, We note that even in the GC only the giant stars
can be presently observed. Rotational broadening in the giant spectra may be
marginally detectable with high resolution spectroscopy, and could bolster the
case for the existence of an underlying very dense population of faint stars.

To summarize, we have shown that MS stars in a substantial volume of the dense
cusps around massive black holes are likely to rotate at a significant fraction
of the centrifugal break-up velocity due to stochastic spin-up by hyperbolic
tidal encounters.

\appendix

\section{Linear tidal coupling coefficients for hyperbolic orbits}

\label{sec:Tl}

Following the formalism of Press \& Teukolsky (\cite{Pre77}), the linear tidal
coupling coefficients are expressed as 

\[
T_{l}(\eta )=2\pi ^{2}\sum _{n}\left| Q_{nl}\right| ^{2}\sum ^{l}_{m=-l}\left| K_{nlm}\right| ^{2}\, ,\]
where \( Q_{nl} \) are overlap integrals that depend only on the stellar structure.
The orbit enters in the term 
\begin{equation}
K_{nlm}=\frac{W_{lm}}{\pi }\left( \frac{\rp ^{3}}{1+\mi }\right) ^{1/2}(1+e)^{-l+1/2}\int _{0}^{\phi _{\mathrm{max}}}d\phi (1+e\cos \phi )^{l-1}\cos (\widetilde{\omega }_{nl}\widetilde{t}(\phi )+m\phi )\, ,
\end{equation}
where \( \widetilde{\omega }_{nl} \) is the frequency of the mode, \( \phi  \)
is the angular position of the impactor in a coordinate system centered on the
target star (\( \phi =0 \) at periastron), and where
\begin{equation}
\phi _{\mathrm{max}}=\arccos \left( \frac{1}{e}\right) \, ,
\end{equation}
and 
\begin{equation}
W_{lm}=(-1)^{\frac{l+m}{2}}\frac{\left( \frac{4\pi }{2l+1}(l-m)!(l+m)!\right) ^{1/2}}{2^{l}\left( \frac{l-m}{2}\right) !\left( \frac{l+m}{2}\right) !}\, .
\end{equation}
The time along the hyperbolic orbit as function of the angle \( \phi  \) is
given by 
\begin{equation}
\widetilde{t}(\phi )=\frac{(1+e)^{3/2}\rp ^{3/2}}{(1+\mi )^{1/2}}\left( \frac{\sin \phi }{1+e\cos \phi }\frac{e}{e^{2}-1}-\left[ e^{2}-1\right] ^{-3/2}\log \left[ \frac{e+\cos \phi +(e^{2}-1)^{1/2}\sin \phi }{1+e\cos \phi }\right] \right) \, .
\end{equation}

Tables~\ref{t:Qwsun} and \ref{t:Qwrg} list the mode frequencies and the overlap
integrals for the solar and giant models that we investigate in this work. The
classification of the giant modes is complicated by the large value of the Brunt-V\"{a}is\"{a}l\"{a} frequency
in the core, compared to the \( f \)-mode frequencies, thereby giving rise
to a mix of \( p \) and \( g \) mode behavior. The corresponding values for
ideal gas \( n=1.5 \), \( 2 \), and \( 3 \) polytropes are given in Lee \&
Ostriker (\cite{Lee86}). 

\acknowledgements

We are grateful to P. Demarque for providing us with the stellar structure model
of the \( \alpha  \)UMa giant.

\pagebreak

\pagebreak

\begin{table}[h]
{\centering \begin{tabular}{cccccccccc}
\hline 
Type&
\( M \) &
\( R \)&
\( V_{e} \)&
\( V_{k} \)&
\( V_{\mathrm{obs}} \)&
\( K \)&
\( T \)&
\( \widetilde{I} \)&
\( \Eb  \)\\
&
\( M_{\odot } \)&
\( R_{\odot } \)&
\( \mathrm{km}\, \mathrm{s}^{-1} \)&
\( \mathrm{km}\, \mathrm{s}^{-1} \)&
\( \mathrm{km}\, \mathrm{s}^{-1} \)&
mag&
yr&
&
\\
\hline 
K2(V)&
0.76&
0.75&
621&
440&
\( \sim 2 \)&
\( \sim 22 \)&
\( >10^{10} \)&
0.07&
1.65\\
G5(III)&
2.4&
8&
340&
240&
\( \sim 5 \)&
\( \sim 16 \)&
\( 1.6\times 10^{8} \)&
0.13&
6.11\\
\( n=1.5 \)&
---&
---&
---&
---&
---&
---&
---&
0.21&
0.85\\
\end{tabular}\par}

\caption{\label{tab: stype}The parameters of representative MS and giant stars (Zombeck
\cite{Zom90}; Lang \cite{Lan91}). The mean observed equatorial circular velocity
\protect\( V_{\mathrm{obs}}\protect \) is from Gray (\cite{Gra92}). Over the
time \protect\( T\protect \), the star is as bright or brighter than the quoted
apparent \protect\( K\protect \) magnitude (for stars in the GC, \protect\( \Delta =14.5^{\mathrm{m}}\protect \),
\protect\( A_{K}=3.5^{\mathrm{m}}\protect \)) and has a radius as large or
larger than the quoted value (based on the twice solar metallicity stellar tracks
of Schaerer et al. \cite{Sch93}). The moment of inertia \protect\( \MOI \protect \),
and the binding energy \protect\( \Eb \protect \) were estimated from stellar
structure models of the Sun (Christensen-Dalsgaard et al. \cite{Chr96}) and
the \protect\( \alpha \mathrm{UMa}\protect \) giant (\protect\( M=4.25\, M_{\odot }\protect \),
\protect\( R=27.4\, R_{\odot }\protect \)) (Guenther et al. \cite{Gue00}).
For comparison, \protect\( \MOI \protect \) and \protect\( \Eb \protect \)
for an ideal gas \protect\( n=1.5\protect \) polytrope are also listed.}
\end{table}

\begin{table}[h]
{\centering \begin{tabular}{ccc}
\hline 
Initial mass range &
Remnant mass &
Mass fraction \\
\( M_{\odot } \)&
\( M_{\odot } \)&
\\
\hline 
0.8--1.5&
0.6&
0.03\\
1.5--2.5&
0.7&
0.08\\
2.5--8&
1.1&
0.12\\
8--30&
1.4&
0.03\\
\textgreater{}30&
10&
0.01\\
\hline 
\end{tabular}\par}

\caption{\label{tab:rem} Stellar remnant mass as function of initial stellar mass (Meylan
\& Mayor \cite{Mey91}; Timmes, Woosley \& Weaver \cite{Tim96}) and its mass
fraction in the continuous star forming stellar population model for the GC
(Alexander \& Sternberg \cite{Ale99b}).}
\end{table}

\begin{table}[h]
{\centering \begin{tabular}{cccc}
\hline 
Type&
Mass range&
\( \alpha  \)&
Mass fraction\\
&
\( M_{\odot } \)&
&
\\
\hline 
MS&
0.4-4.0&
2.35&
0.73\\
WD&
0.7-1.1&
0.0&
0.23\\
NS&
1.2-1.5&
2.35&
0.03\\
BH&
7.0-12.0&
2.35&
0.01\\
\hline 
\end{tabular}\par}

\caption{\label{tab:PMF}Model for the impactor mass function in the inner GC. A \protect\( df/dm\propto m^{-\alpha }\protect \)
mass distribution is assumed within the mass range.}
\end{table}

\begin{table}[h]
{\centering \begin{tabular}{ccccccccc}
\hline 
\( l \)&
\multicolumn{2}{c}{ 2}&
\multicolumn{2}{c}{ 3}&
\multicolumn{2}{c}{ 4}&
\multicolumn{2}{c}{ 5}\\
\hline 
Mode&
\( \widetilde{w}^{2}_{nl} \)&
\( \left| Q_{nl}\right|  \)&
\( \widetilde{w}^{2}_{nl} \)&
\( \left| Q_{nl}\right|  \)&
\( \widetilde{w}^{2}_{nl} \)&
\( \left| Q_{nl}\right|  \)&
\( \widetilde{w}^{2}_{nl} \)&
\( \left| Q_{nl}\right|  \)\\
\hline 
\( p_{6} \)&
1.28(+2)&
2.38(-2)&
1.39(+2)&
1.12(-2)&
1.49(+2)&
5.39(-3)&
1.59(+2)&
2.74(-3)\\
\( p_{5} \)&
9.70(+1)&
3.26(-2)&
1.06(+2)&
1.58(-2)&
1.14(+2)&
8.00(-3)&
1.23(+2)&
4.28(-3)\\
\( p_{4} \)&
7.02(+1)&
4.67(-2)&
7.76(+1)&
2.34(-2)&
8.48(+1)&
1.22(-2)&
9.18(+1)&
6.82(-3)\\
\( p_{3} \)&
4.79(+1)&
6.85(-2)&
5.36(+1)&
3.48(-2)&
5.90(+1)&
1.91(-2)&
6.45(+1)&
1.14(-2)\\
\( p_{2} \)&
2.95(+1)&
1.06(-1)&
3.33(+1)&
5.61(-2)&
3.72(+1)&
3.32(-2)&
4.12(+1)&
2.16(-2)\\
\( p_{1} \)&
1.70(+1)&
1.47(-1)&
1.82(+1)&
9.18(-2)&
2.01(+1)&
6.36(-2)&
2.23(+1)&
4.66(-2)\\
\( f \)&
1.39(+1) &
1.39(-1)&
1.65(+1)&
5.54(-2)&
1.77(+1)&
8.62(-3)&
1.84(+1)&
7.56(-4)\\
\( g_{1} \)&
9.12(+0)&
4.58(-2)&
1.17(+1)&
8.47(-3)&
1.35(+1)&
4.80(-3)&
1.47(+1)&
2.40(-3)\\
\( g_{2} \)&
6.92(+0)&
1.13(-1)&
8.94(+0)&
4.46(-2)&
1.08(+1)&
1.64(-2)&
1.23(+1)&
7.29(-3)\\
\( g_{3} \)&
5.11(+0)&
1.07(-1)&
7.02(+0)&
8.99(-2)&
8.58(+0)&
3.81(-2)&
1.00(+1)&
1.56(-2)\\
\( g_{4} \)&
3.84(+0)&
8.31(-2)&
5.81(+0)&
1.10(-1)&
7.13(+0)&
1.00(-1)&
8.35(+0)&
4.85(-2)\\
\( g_{5} \)&
2.92(+0)&
5.66(-2)&
4.75(+0)&
7.20(-2)&
6.35(+0)&
1.04(-1)&
7.51(+0)&
1.29(-1)\\
\( g_{6} \)&
2.31(+0)&
3.91(-2)&
3.84(+0)&
4.36(-2)&
5.38(+0)&
4.76(-2)&
6.76(+0)&
5.20(-2)\\
\( g_{7} \)&
1.82(+0)&
2.77(-2)&
3.13(+0)&
2.90(-2)&
4.45(+0)&
2.73(-2)&
5.76(+0)&
2.31(-2)\\
\( g_{8} \)&
1.49(+0)&
1.99(-2)&
2.59(+0)&
2.04(-2)&
3.76(+0)&
1.84(-2)&
4.93(+0)&
1.48(-2)\\
\( g_{9} \)&
1.23(+0)&
1.44(-2)&
2.19(+0)&
1.47(-2)&
3.20(+0)&
1.30(-2)&
4.24(+0)&
1.03(-2)\\
\( g_{10} \)&
1.04(+0)&
1.07(-2)&
1.85(+0)&
1.12(-2)&
2.76(+0)&
1.02(-2)&
3.69(+0)&
8.20(-3)\\
\hline 
\end{tabular}\par}

\caption{\label{t:Qwsun} Mode frequencies and overlap integrals for the solar model
of Christensen-Dalsgaard et al. (\cite{Chr96}).}
\end{table}

\begin{table}[h]
{\centering \begin{tabular}{ccccccc}
\hline 
\( l \)&
\multicolumn{2}{c}{ 2}&
\multicolumn{2}{c}{ 3}&
\multicolumn{2}{c}{ 4}\\
\hline 
Mode&
\( \widetilde{w}^{2}_{nl} \)&
\( \left| Q_{nl}\right|  \)&
\( \widetilde{w}^{2}_{nl} \)&
\( \left| Q_{nl}\right|  \)&
\( \widetilde{w}^{2}_{nl} \)&
\( \left| Q_{nl}\right|  \)\\
\hline 
\( p_{8} \)&
4.28(+1)&
1.25(-2)&
4.53(+1)&
1.62(-2)&
4.83(+1)&
1.86(-2)\\
\( p_{7} \)&
3.66(+1)&
1.66(-2)&
3.40(+1)&
2.82(-2)&
3.67(+1)&
3.18(-2) \\
\( p_{6} \)&
3.08(+1)&
2.29(-2)&
2.21(+1)&
6.09(-2)&
3.15(+1)&
4.12(-2) \\
\( p_{5} \)&
2.19(+1)&
4.35(-2)&
1.60(+1)&
1.02(-1)&
2.62(+1)&
5.86(-2) \\
\( p_{4} \)&
1.45(+1)&
8.40(-2)&
1.11(+1)&
1.66(-1)&
1.68(+1)&
1.13(-1) \\
\( p_{3} \)&
1.27(+1)&
1.05(-1)&
9.24(+0)&
2.47(-1)&
1.01(+1)&
2.27(-1) \\
\( p_{2} \)&
8.12(+0)&
2.36(-1)&
6.55(+0)&
1.26(-1)&
6.81(+0)&
3.59(-2) \\
\( p_{1} \)&
6.10(+0)&
2.68(-1)&
4.45(+0)&
9.06(-2)&
5.02(+0)&
3.42(-2) \\
\( f \)&
3.57(+0)&
1.84(-1)&
3.39(+0)&
8.86(-2)&
2.19(+0)&
3.69(-2) \\
\end{tabular}\par}

\caption{\label{t:Qwrg} Mode frequencies and overlap integrals for the giant model
of Guenther et al. (\cite{Gue00}).}
\end{table}

\end{document}